\documentclass[12pt]{article}
\usepackage{epsfig,amsfonts,amssymb}
\usepackage{hyperref}
\usepackage{cite}
\input epsf.sty
\usepackage{cite}

\def\ZZZ{{\hbox{ Z\kern-1.6mm Z}}}
\def\RRR{{\hbox{ R\kern-2.4mm R}}}
\def\CCC{{\hbox{ C\kern-2.0mm C}}}
\def\zzz{{\hbox{z\kern-1mm z}}}

\newcommand{\nn}{\nonumber \\}

\newcommand{\vt}{\vartheta}

\newcommand{\qeq}{{\hbox{=\kern-2.3mm ? \kern.5mm }}}
\renewcommand{\qeq}{=}

\newcommand{\II}{{\cal I}}

\newcommand{\wt}{\widetilde}
\newcommand{\wh}{\widehat}

\newcommand{\NN}{{\cal N}}

\newcommand{\be}{\begin{equation}}
\newcommand{\ee}{\end{equation}}
\newcommand{\ben}{\begin{eqnarray}\displaystyle}
\newcommand{\een}{\end{eqnarray}}

\newcommand{\refb}[1]{(\ref{#1})}

\newcommand{\sectiono}[1]{\section{#1}\setcounter{equation}{0}}

\def\one{{\hbox{ 1\kern-.8mm l}}}
\def\zero{{\hbox{ 0\kern-1.5mm 0}}}

\begin{document}

\baselineskip 24pt

\begin{center}
{\Large \bf  Arithmetic of $\NN=8$ Black Holes}

\end{center}

\vskip .6cm
\medskip

\vspace*{4.0ex}

\baselineskip=18pt

\centerline{\large \rm   Ashoke Sen }

\vspace*{4.0ex}

\centerline{\large \it Harish-Chandra Research Institute}

\centerline{\large \it  Chhatnag Road, Jhusi,
Allahabad 211019, INDIA}

\vspace*{1.0ex}
\centerline{E-mail:  sen@mri.ernet.in, ashokesen1999@gmail.com}

\vspace*{5.0ex}

\centerline{\bf Abstract} \bigskip

The microscopic formula for the degeneracies of 1/8 BPS black
holes in type II string theory compactified on a six dimensional
torus can be expressed as a sum of several
terms. One of the terms is a
function of the Cremmer-Julia invariant and gives 
the leading contribution
to the entropy in the large charge limit. The other terms, which give
exponentially subleading contribution,
depend not
only on the Cremmer-Julia invariant, but also on the arithmetic
properties of the charges, and in fact exist only when the charges
satisfy special arithmetic properties. 
We identify the origin of these terms in the macroscopic formula
for the black hole entropy, based on quantum entropy function,
as the contribution from non-trivial
saddle point(s) in the
path integral of string theory over the near horizon geometry.
These saddle points 
exist only when the charge vectors satisfy the 
arithmetic properties required
for the corresponding term in the
microscopic formula to exist. Furthermore the leading
contribution from these saddle points in the large charge limit
agrees with the leading asymptotic behaviour of the corresponding term
in the degeneracy formula. 

\vfill \eject

\baselineskip=18pt

\tableofcontents

\sectiono{Introduction}  \label{s1}

We now have a good understanding of the spectrum of 
BPS dyons in the
four dimensional
$\NN=8$ supersymmetric string theory obtained by compactifying
type II string theory on  $T^6$. 
In particular the exact degeneracies are known for a class of
1/8 BPS dyons in this 
theory\cite{9903163,0506151,0506228,0508174,0803.1014,0804.0651}. 
On the other hand 
after taking into account the effect of gravitational backreaction
these dyons are expected to become black holes with finite area
event horizon and hence have finite macroscopic entropy. In the
limit of large charges the Bekenstein-Hawking entropy of these
black holes match the statistical entropy, --
logarithm of the microscopic degeneracy of states carrying the
same charges. 

In a recent series of papers an algorithm for computing
the exact macroscopic
degeneracy of extremal black holes was 
proposed\cite{0805.0095,0809.3304,0903.1477}. This
algorithm -- known as the quantum entropy function -- equates
the macroscopic degeneracy with the partition function 
of string theory
in the euclidean
near horizon geometry of the black hole that contains
an $AdS_2$ factor.\footnote{This gives the contribution from a
single centered black hole horizon. The contribution to the
degeneracy from a generic multi-centered black hole is obtained
by taking the product of the contribution from each horizon and
the degeneracies due to the hair modes, -- degrees of freedom
living outside the horizon\cite{0903.1477}.
For 1/8 BPS black holes in $\NN=8$
supersymmetric string theories we expect the effect of multi-centered
configurations to be absent for the charges of interest
to us -- both for the wall 
crossing\cite{0803.1014} and
also for the total index\cite{0903.2481}.
Hence we can concentrate on the contribution
from single centered black holes represented by a single
$AdS_2$ factor.} 
More precisely the path integral contributing to the partition
function includes
sum over all configurations which asymptotes to the near horizon
geometry of the black hole near the boundary of $AdS_2$.
Given this algorithm for computing the exact macroscopic degeneracies
it is natural to compare this with the exact microscopic degeneracies
which are known.\footnote{Since on the macroscopic side we
compute degeneracies but on the microscopic side we compute an
appropriate helicity trace index\cite{9708062,9708130}
 -- the 14th helicity trace $B_{14}$ in the
present example -- one might wonder if it is appropriate to compare
the two. It was argued in \cite{0903.1477} 
that as long as the only hair degrees
of freedom of the black hole are the fermion zero modes associated
with broken supersymmetries -- a condition we expect to be satisfied
for the configurations involving only D-brane charges as studied 
here\cite{0903.1477} -- the macroscopic degeneracy should
agree with the microscopic index $-B_{14}$. For this reason
we shall often refer to $-B_{14}$ as the microscopic degeneracy,
and the various microscopic formul\ae\ given in the text refer
to this index.}

A full analysis will be beyond the scope of the present work as it
would require explicit evaluation of the path integral of string
theory in appropiate backgrounds. Instead in this paper we 
investigate a particular aspect of this problem. The microscopic
result for the degeneracy can be expressed as a sum over
finite number of
terms. One of these terms, which gives the leading contribution to the
entropy in the large charge limit, is a function of the Cremmer-Julia
invariant constructed out of the charges\cite{cremmer,9602014}.
The other terms, which give exponentially suppressed contributions in the
large charge limit, exist only when the charge vectors satisfy special
arithmetic properties. We show that for each of these terms, one can
identify saddle points, constructed as appropriate orbifolds 
(without fixed points) of the
near horizon geometry of the black hole, satisfying the following
properties:
\begin{enumerate}
\item The geometry associated with the saddle point coincides with
the near horizon geometry of the black hole near the boundary of
$AdS_2$. This shows that this is a valid configuration to be included
in the path integral over the string fields.
\item  The saddle point exists only when the charge vectors
satisfy special arithemetic properties -- the same properties for
which the corresponding term in the microscopic
degeneracy formula exists.
\item For large charges 
the contribution to the path integral from this saddle point
has the same behaviour as the corresponding term in the degeneracy
formula.
\end{enumerate}
Thus these saddle points are the ideal candidates for representing the
corresponding terms in the degeneracy formula. This generalizes
similar results for $\NN=4$ supersymmetric string 
theories\cite{0903.1477} 
(see also {\cite{0810.3472,0904.4253}).

\sectiono{$N=8$ Dyon Spectrum: Microscopic Results} \label{s2}

We consider type IIB string theory on $T^6$, which we shall label
as $T^4\times S^1\times \wt S^1$ by regarding two circles inside
$T^6$ as special. In this theory we consider a
system of $D5/D3/D1$ branes wrapped on $4/2/0$ cycles of $T^4$
times either $S^1$ or $\wt S^1$. We shall denote the 
charges of the D-branes 
wrapped on
$S^1$ by $P$ and the charges of
the D-branes wrapped on $\wt S^1$
by $Q$. Both $Q$ and $P$ are 8 dimensional vectors reflecting
the dimension of the even cohomology of $T^4$. A perturbatively
realized symmetry which acts within this class of charges is
$SO(4,4;\ZZZ)\times SL(2,\ZZZ)$ where the former is associated
with the duality symmetries of $T^4$ and the latter is associated
with the global diffeomorphism symmetry of $S^1\times \wt S^1$.
A U-duality transformation maps this system to type IIA string theory
on $T^6$ where all charges arise in the NSNS sector, with $Q$
representing the electric charges and $P$ representing the magnetic
charges.\footnote{This can be done by first making a 
T-duality transformation on the circle $\wt S^1$ and then making
a $\ZZZ_2$ U-duality transformation
that maps $(-1)^{F_L}$
to a geometric symmetry $\II_4$ that reverses the signs of
all the coordinates of $T^4$ and vice-versa\cite{9508064}.
Thus the $(-1)^{F_L}$  odd gauge fields from the RR sector
are mapped to $\II_4$ odd gauge
fields given by the  dimensional reduction of the metric and the
NSNS 2-form fields along $T^4$.
 \label{f1}}
In this case $SO(4,4;\ZZZ)$ appears as a subgroup
of the T-duality group  and 
$SL(2,\ZZZ)$ appears as the S-duality group
of the theory. In view of
this we shall call $SO(4,4;\ZZZ)$ and $SL(2,\ZZZ)$ as T- and S-duality
sysmmetries even though in the description we are using both are
part of the T-duality group.

The intersection matrix $L$
of the even homology cycles of
$T^4$ defines a natural inner product
between the charge vectors $Q$ and $P$:
\be \label{einner}
Q^2\equiv Q^T L Q, \quad P^2 \equiv P^TLP, \quad Q\cdot P
\equiv Q^T L P\, .
\ee
 We  define
\be \label{e1}
\ell_1 = \gcd\{Q_i P_j - Q_j P_i\}, \qquad
\ell_2 = \gcd\left( {Q^2\over 2}, {P^2\over 2}, Q\cdot P\right) ,
\ee
where $Q_i$ and $P_i$ are the components of $Q$ and $P$ in some
primitive basis of the charge lattice $\Lambda$. 
$\ell_1$ and $\ell_2$ remain invariant under S- and T-duality
transformations of $(Q,P)$.
The formula for the dyon
degeneracy carrying charges $(Q,P)$ is known  in the 
case\footnote{It may be possible to relax this condition by using the
results of \cite{9903163} for non-primitive charge vectors.}
\be \label{e2}
\gcd(\ell_1, \ell_2)=1\, .
\ee
In this case the degeneracy formula
for the charge vectors considered here takes the 
form
\be \label{e3}
d(Q,P) = (-1)^{Q\cdot P+1}
\sum_{s|\ell_1 \ell_2} \, s \, \wh c( \Delta(Q,P)/s^2)\, ,
\ee
where $\Delta(Q,P)$ is the Cremmer-Julia 
invariant\cite{cremmer,9602014}
\be \label{e4}
\Delta(Q,P) = Q^2 P^2 - (Q\cdot P)^2\, ,
\ee
and $\wh c(u)$ is defined through the 
relations\cite{9903163,0506151}
\be \label{ek6.5}
-\vt_1(z|\tau)^2 \, \eta(\tau)^{-6} \equiv \sum_{k,l} \wh c(4k-l^2)\, 
e^{2\pi i (k\tau+l z)}\, .
\ee
$\vt_1(z|\tau)$ and $\eta(\tau)$ are respectively the odd Jacobi
theta function and the Dedekind eta function.  The derivation of
\refb{e2}, \refb{e3} 
has been reviewed in appendix \ref{sa}.

For large charges we have
\be \label{e5}
\wh c (\Delta) \sim (-1)^{\Delta +1}\, 
\Delta^{-2}\, \exp(\pi\sqrt{\Delta})\, .
\ee
Thus the $s$-th term in the sum grows as $\exp(\pi\sqrt{\Delta}/s)$.
In this limit the $s=1$ term dominates, and the
contribution to the entropy reduces to the  Bekenstein-Hawking
entropy of the black hole given by $\pi\sqrt{\Delta}$. However the
terms with $s>1$ are significant in that they appear only when the
charge vectors satisfy some special arithmetic properties. Thus
one should be able to detect the origin of these terms in the
macroscopic description by identifying contributions which appear
only when the charge vectors satisfy these special
arithmetic properties.\footnote{Different aspects of the relationship
between arithmetic and black holes have been studied in
\cite{9807056,9807087,0401049}.}

\sectiono{$N=8$ Dyon Spectrum: Macroscopic Viewpoint}

According to the proposal of \cite{0809.3304,0903.1477}, 
the macroscopic entropy
of an extremal black hole is given by the result of
path integral over geometries whose asymptotic form coincide
with the near horizon geometry of the black hole. In the
case under consideration the near horizon metric of the
Euclidean black hole carrying charges $(Q,P)$ takes the
form:
\be \label{ech3}
ds^2 = v \, \left( {dr^2\over r^2-1}+(r^2-1)\,  d\theta^2\right)
+ w (d\psi^2 + \sin^2\psi d\phi^2) + {R^2\over\tau_2} 
\left|dx^4+\tau dx^5
\right|^2 + \sum_{m,n=6}^9 \, \wh g_{mn}  dx^m dx^n\, , 
\ee
where $v$, $w$, $R$  are real constants,
$\tau=\tau_1+i\tau_2$ is a complex constant,
and $\wh g_{mn}$ are real constants labelling
the metric 
along $T^4$. 
$(r,\theta)$ label an Euclidean $AdS_2$ space,
$(\psi,\phi)$ label a 2-sphere, $x^4$ and $x^5$ label the
coordinates along $\wt S^1$ and $S^1$ respectively and
$x^6,x^7,x^8,x^9$ are coordinates along $T^4$. 
Each of the coordinates $x^4,\cdots x^9,\theta,\phi$ has period
$2\pi$.
The background also
contains constant values of various scalar fields and components
of $p$-form fields along $T^4\times S^1\times \wt S^1$,
and fluxes of various RR fields. In the six dimensional
description, in which all the RR field strengths can be regarded
as self-dual or anti-self-dual 
3-forms after dimensional reduction on $T^4$,
$Q$ represents
RR fluxes through the 3-cycle spanned by $(x^5,\psi,\phi)$ and
$P$ represents
RR fluxes through the 3-cycle spanned by $(x^4,\psi,\phi)$.
The (anti-)self-duality constraints on the 
RR field strengths in six dimensions
relate the fluxes through the $(x^4,r,\theta)$ and
$(x^5,r,\theta)$ planes to those through the $(x^5,\psi,\phi)$
and $(x^4,\psi,\phi)$ planes. The charges $(Q,P)$ also determine,
up to flat directions, the parameters $v,w,R,\tau,\wh g_{mn}$
and the background values of various scalars and $p$-form fields.

In our analysis we shall assume that the $SO(4,4;\ZZZ)_T\times 
SL(2,\ZZZ)_S$ symmetry is a symmetry of 
string theory in the near horizon geometry,
\i.e.\ two different configurations which differ from each other by the
action of this symmetry on $(Q,P)$ give identical results for the
partition function. This assumption is natural since both
of these are perturbative symmetries of the theory in the
description in which we are working.
We can then make use of these duality symmetries to bring the
charge vectors $(Q,P)$ to a specific form and carry out the
analysis; the result for
a general charge vector can be recovered by making an
appropriate $SO(4,4;\ZZZ)_T \times SL(2,\ZZZ)_S$ 
duality transformation.\footnote{More generally we can assume that
the full perturbative duality symmetry $SO(6,6;\ZZZ)$ is a symmetry
of string theory in the near horizon geometry. In that case our results
extend to a more general configuration of D-branes which can be related
to
the configurations analyzed here by an $SO(6;6;\ZZZ)$ transformation.
A general configuration of D-branes wrapped on various cycles of
$T^6$ is characterized by a 32 dimensional charge vector
transforming in the spinor representation of $SO(6,6;\ZZZ)$, but we
do not know under what condition on this charge vector it can be
related to a configuration analyzed here by an 
$SO(6,6;\ZZZ)$ transformation.}
Now it was shown in \cite{0801.0149} that 
given any pair of charge vectors
$(Q,P)$ we can use S-duality transformations
to bring it to the form
\be \label{e6}
(Q,P) = (\ell_1 Q_0, P_0), \qquad \gcd\{Q_{0i}P_{0j}
- Q_{0j} P_{0i}\}=1\, ,
\ee
where $\ell_1$ has been defined in \refb{e1}
and $Q_0$ and $P_0$ are elements of the charge lattice
$\Lambda$. We shall
use this representation.
Furthermore it follows from the analysis of
\cite{0712.0043} that if for a charge vector 
of the form \refb{e6}  we define
\be \label{e12}
n=Q_0^2 / 2, \quad m=P_0^2 / 2, \quad p = Q_0\cdot P_0 \, ,
\ee
then with the help of T-duality trasnformations we can bring
$Q_0$ and $P_0$ inside a four dimensional subspace of the
full eight dimensional $SO(4,4)$ lattice, and label them
by the vectors
\be \label{e13}
Q_0 = \pmatrix{1 \cr n\cr 0 \cr 0}, \qquad
P_0 = \pmatrix{0 \cr p\cr 1 \cr m}\, ,
\ee
the metric in this subspace being given by
\be \label{e14}
L = \pmatrix{0 & 1 & 0 & 0\cr 1 & 0 & 0 & 0\cr
0 & 0 & 0 & 1\cr 0 & 0 & 1 & 0}\, .
\ee
Furthermore we can choose the embedding of this four dimensional
subspace in the even cohomology of $T^4$ such that the four rows
of $Q=\ell_1 Q_0$
represent the RR 5-form fluxes through the $\psi\phi x^5x^6x^7$,
$\psi\phi x^5x^8x^9$, $\psi\phi x^5x^6x^8$ and 
$\psi\phi x^5x^9x^7$ cycles respectively,
and the four rows of $P=P_0$  
represent the RR 5-form fluxes through the $\psi\phi x^4x^6x^7$,
$\psi\phi x^4x^8x^9$, $\psi\phi x^4x^6x^8$ and 
$\psi\phi x^4x^9x^7$ cycles respectively. 
Thus the RR 5-form flux is of the form
\ben \label{ey1}
F &=& {1\over 32\pi^4} \, \sin\psi \, d\psi\wedge d\phi \,
\bigg[ \ell_1 \, dx^5 \wedge dx^6\wedge dx^7 
+ \ell_1 \, n\, dx^5 \wedge dx^8\wedge dx^9
+ p \, dx^4 \wedge dx^8\wedge dx^9 \nn
&& \qquad 
+ dx^4 \wedge dx^6\wedge dx^8
- m\, dx^4 \wedge dx^7\wedge dx^9\bigg]\, ,
\een
where we have normalized the flux so that its integral over
any 5-cycle is an integer. {}From the definition
of $\ell_2$ given in \refb{e1}, and \refb{e2}, \refb{e6},
\refb{e12} it follows that
\be \label{e14.5}
\ell_2 = \gcd(m,n,p), \qquad \gcd(m, \ell_1)=1\, .
\ee
The requirement of self-duality also forces us to have RR 5-form
flux through $AdS_2$ times appropriate 3-cycles of $T^6$.
They can be determined from \refb{ey1} but we shall not write them
down explicitly.

We are now ready to describe our proposal for the macroscopic
origin of the different terms appearing in the microscopic
formula \refb{e3}. 
Due to the condition \refb{e2}, any
$s$ contributing to the sum in \refb{e3} must have the form
\be \label{e7}
s=s_1\, s_2, \quad s_1,s_2\in\ZZZ,
\quad s_1|\ell_1, \quad s_2|\ell_2, \quad \gcd(s_1,s_2)=1
\, .
\ee
We  propose that the $s$'th term in the sum in \refb{e3}
arises from the orbifold of the geometry \refb{ech3} by a
$\ZZZ_{s_1}\times \ZZZ_{s_2}$ transformation. 
The $\ZZZ_{s_1}$
is generated by
\be \label{ech4}
(\theta,\phi,x^5)\to \left(\theta+{2\pi\over s_1},\phi+{2\pi 
\over s_1},x^5
+{2\pi k_1\over s_1}
\right)\, , \quad k_1\in \ZZZ, \quad \gcd(s_1,k_1)=1\, .
\ee
On the other hand the $\ZZZ_{s_2}$ action is generated by
\be \label{e8}
(\theta,\phi,x^9 )\to \left(\theta+{2\pi\over s_2},\phi+{2\pi 
\over s_2}, x^9 + {2\pi k_2\over s_2}
\right)\, , \quad k_2\in\ZZZ, \quad \gcd(s_2, k_2) = 1\,  .
\ee
Since $s_1$ and $s_2$ do not have any common factor, one can
also regard this as a $\ZZZ_s$ orbifold generated by
\ben \label{e8.5}
&&(\theta,\phi,x^5, x^9)\to \left(\theta+{2\pi\over s },\phi+{2\pi 
\over s },x^5
+{2\pi j_1\over s_1}, x^9 + {2\pi j_2\over s_2}
\right)\, , \nn
&& \qquad  j_1,j_2 \in \ZZZ, \quad \gcd(j_1,s_1 )=
\gcd (j_2, s_2) =1\, .
\een
Since $x^5$ and $x^9$ circles are non-contractible, $\ZZZ_s$ acts
freely and hence this orbifold does not have any fixed point.

A simpler description of this orbifold can be given by
introducing new coordinates
\be \label{ey2}
\xi= \alpha \, x^5 + \beta \, x^9, \qquad
\eta = -s_1 \wt j_2 \, x^5 + s_2 \, \wt j_1 \, x^9,
\qquad \wt j_i \equiv j_i /k, \quad k\equiv \gcd(j_1,j_2)\, ,
\ee
where $(\alpha, \beta)$ are chosen such that
\be \label{ey3}
\alpha \, s_2 \, \wt j_1 + \beta \, s_1\, \wt j_2 =1, 
\qquad \alpha, \beta \in \ZZZ\, .
\ee
This is possible since $\gcd(s_2 \, \wt j_1, s_1 \wt j_2)=1$
by construction. 
Since the transformation \refb{ey2} is unimodular, $\xi$
and $\eta$ are both periodic coordinates with period $2\pi$.
In terms of these new coordinates $(\xi,\eta)$
we can express the 
five form flux \refb{ey1} and the orbifold action \refb{e8.5} as
\ben \label{ey6}
F &=& {1\over 32\pi^4} \, \sin\psi \, d\psi\wedge d\phi \,
\bigg[ \ell_1 \, (s_2 \wt j_1 d\xi - \beta d\eta) 
\wedge dx^6\wedge dx^7 
- \ell_1 \, n\, dx^8\wedge d\xi\wedge d\eta \nn
&& \qquad \qquad
+ p \, dx^4 \wedge dx^8\wedge (s_1 \wt j_2  \, d\xi + \alpha d\eta)
+ dx^4 \wedge dx^6\wedge dx^8 \nn && \qquad \qquad 
- m\, dx^4 \wedge dx^7\wedge (s_1 \wt j_2  \, d\xi + \alpha d\eta)\bigg]
\, , 
\een
\ben \label{ey5}
&&(\theta,\phi,\xi,\eta)\to \left(\theta+{2\pi\over s },\phi+{2\pi 
\over s },\xi
+{2\pi k \over s}, \eta
\right) \, .
\een
It also follows from the definition of $k$ given in \refb{ey2} that
$\gcd(k,s)=1$.

We are now ready to test the consistency of this orbifold.
Since at the origin $r=1$ of $AdS_2$ the $\theta$ translation has
no effect,
the effect of taking the $\ZZZ_s$ orbifold \refb{ey5} is to
reduce the flux through any 5-cycle sitting at $r=1$
and containing $(\psi,\phi,\xi)$
to $1/s$ times its original value. Thus in order that the orbifold
satisfies the flux quantization laws, the coefficient of every term inside
the square bracket in \refb{ey6}, containing 
$d \xi$, must be an integer multiple of
$s=s_1s_2$. Examining \refb{ey6} and using 
the fact that $\ell_1$ is divisible by $s_1$ and $m,n,p$ are
divisible by $s_2$ due to \refb{e14.5}
we see that this is indeed the case.
Thus \refb{ey5} and hence \refb{e8.5} describes a
consistent orbifold in string theory.
Conversely, unless $s_1|\ell_1$ and $s_2|\ell_2$,  the
coefficient of one of the terms containing $d\xi$ inside $[~]$ 
will fail to be
divisible by $s$ and hence the orbifold \refb{e8.5}
will not satisfy
flux quantization rule.

Following the procedure of \cite{0810.3472,0903.1477,0904.4253} 
one can show that
\begin{enumerate}
\item These orbifolds have the same asymptotic behaviour as the
near horizon geometry \refb{ech3}, and hence must be included in
the path integral for computing the quantum entropy function.
\item The classical contribution to the path integral from these
orbifolds is given by
\be \label{e9}
\exp(\pi \sqrt{\Delta(Q,P)}/s)\, ,
\ee
in agreement with the asymptotic behaviour of the $s$-th term
in the sum in \refb{e3}.
\end{enumerate}
Since a detailed analysis can be found in \cite{0903.1477} 
(section 6) we shall
not repeat it here. One can also show that these orbifolds
preserve the necessary amount of supersymmetry so that
integration over the fermion zero modes associated with the broken
supersymmetry generators does not make the path integral
vanish automatically\cite{0905.2686}.
Thus the contribution to the quantum entropy function
from these orbifolds is the ideal candidate for
reproducing the $s$-th term in
the microscopic formula given in \refb{e3}.

\medskip

{\bf Acknowledgement:} 
I wish to thank Atish Dabholkar and Edward Witten for useful
discussions.
I would like to acknowledge the hospitality of 
LPTHE, Paris
where part of this work was performed.
This work
was supported by the project 11-R\& D-HRI-5.02-0304 and the J.C.Bose
fellowship of the Department of Science and Technology, India.

\appendix

\sectiono{The Dyon Degeneracy Formula}  \label{sa}

In this appendix we shall give a derivation of the dyon degeneracy
formula given in \refb{e2}, \refb{e3} from the duality covariant formula
described in \cite{0804.0651}. Let us consider the 
configurations considered in the text, labelled by the charges
$(Q,P)$.
We shall use the symbol $q$ to denote the pair $(Q,P)$.
Working in the duality frame described in footnote \ref{f1}
one finds that
the two discrete duality invariants $\psi(q)$ and $\chi(q)$ introduced
in \cite{0804.0651} take the 
form:\footnote{Derivation of \refb{ebb1} and
\refb{ebb2} can be found in \cite{0804.0651}.
In a general U-duality frame $q$ belongs 
to the {\bf 56}
representation of the U-duality group $E_{7(7)}(\ZZZ)$,
$\psi(q)$ is the gcd of all the
components of the {\bf 133} representation
constructed from the bilinear $q_a q_b$, 
$\wt q_a$ represents the vector in the
{\bf 56} representation constructed from the trilinear
$q_a q_b q_c$, and
$\chi(q)=\gcd\{q_a \wt q_b - q_b \wt q_a\}$.}
\be \label{ebb1}
\psi(q) = \gcd\left( {Q^2\over 2}, {P^2\over 2}, Q\cdot P,
\{Q_i P_j - Q_j P_i\}\right), 
\ee
and 
\be \label{ebb2}
\chi(q) = \gcd\{ q_a \wt q_b - \wt q_a q_b\},
\qquad \wt q\equiv (\wt Q, \wt P) =
\left(Q^2 P - (Q\cdot P) Q, - P^2 Q + (Q\cdot P) P\right)\, .
\ee
The degeneracy formula given in \cite{0804.0651} is valid for
charge vectors with $\psi(q)=1$ and takes the form
\be \label{ebb4}
d(q)=(-1)^{Q\cdot P+1}
\sum_{s\in\zzz, 2s|\chi(q)} s\, \wh c(\Delta(Q,P)/s^2)\, .
\ee
A detailed derivation of this formula has been given in
\cite{0803.1014,0804.0651} based on earlier 
work\cite{0506151,
0506228} and a recent discussion can be found in
\cite{0903.5517}. 

We shall now show that the restriction $\psi(q)=1$ and
\refb{ebb4} lead to \refb{e2}, \refb{e3}.
{}From  \refb{ebb1} and
 \refb{e1} it follows that
\be \label{ebb3}
\psi(q) =\gcd (\ell_1, \ell_2)\, ,
\ee
and hence the restriction $\psi(q)=1$ reduces to $\gcd(\ell_1,
\ell_2)=1$ as given in \refb{e2}.
In order to show that \refb{ebb4}
reduces to \refb{e3} we need to show that
the condition $2s|\chi(q)$ corresponds to the restriction
$s|\ell_1\ell_2$ as appears in the sum in \refb{e3}. For this we expand
\refb{ebb2}:
\be \label{ebb5}
\chi(q) =
\gcd\left\{ Q^2 (Q_i P_j - Q_j P_i), \quad
P^2 (Q_i P_j - Q_j P_i), \quad
(-P^2 Q_i Q_j - Q^2 P_i P_j + 2\, Q\cdot P\, Q_i P_j)
\right\}\, .
\ee
Now in computing the gcd we can certainly add to the list inside
$\{~\}$ a term that is obtained by antisymmetrizing the last 
set of terms
in the indices $i$ and $j$. This gives
\ben \label{ebb6}
\chi(q) &=&
\gcd\left\{ Q^2 (Q_i P_j - Q_j P_i), \quad
P^2 (Q_i P_j - Q_j P_i), \quad 2 Q\cdot P (Q_i P_j - Q_j P_i),
\right.
\nn
&& \left. \qquad \qquad 
(-P^2 Q_i Q_j - Q^2 P_i P_j + 2\, Q\cdot P\, Q_i P_j)
\right\}\, .
\een
{}From this we see that a necessary 
condition for $2s|\chi(q)$ is
\ben \label{ebb7}
2s&|&\gcd\left\{ Q^2 (Q_i P_j - Q_j P_i), \quad
P^2 (Q_i P_j - Q_j P_i), \quad 2 Q\cdot P (Q_i P_j - Q_j P_i)
\right\} \nn
&=& \gcd(Q^2, P^2,2 Q\cdot P) \times \gcd\{Q_iP_j-Q_j P_i\}
= 2 \ell_1 \ell_2\, ,
\een
and hence
\be \label{ebb8}
s|\ell_1\ell_2\, 
\ee
as given in \refb{e3}. However we also need to show that this
condition is sufficient \i.e.\ that
once \refb{ebb8} is satisfied then $2s$ automatically divides the
last set of terms inside the list in \refb{ebb6}. 
These terms may
be written in two different forms:
\be  \label{ebb9}
\{(-P^2 Q_i Q_j - Q^2 P_i P_j + 2\, Q\cdot P\, Q_i P_j)\}
=\{ -(Q_k P_i - P_k Q_i) \, (Q_k P_j - Q_j P_k)
- Q\cdot P\, (P_i Q_j - Q_i P_j)\}\, .
\ee
The form given in the left hand side 
shows that \refb{ebb9} 
is divisible
by $2\ell_2$ since $2\ell_2=\gcd(P^2,Q^2,2Q\cdot P)$.
On the other hand the form given on the right hand side shows
that it is divisible by $\ell_1$ since $\ell_1
=\gcd\{P_i Q_j - Q_i P_j\}$. 
Now if $\ell_1$ is odd, then it follows from  \refb{e2}
that $\gcd(2\ell_2, \ell_1)=1$
and hence \refb{ebb9}, being divisible by $\ell_1$ and $2\ell_2$,
must be divisible by $2\ell_1\ell_2$. On the other hand if $\ell_1$
is even then  $Q_kP_i-P_k Q_i$ must be
even for every $i,k$ and hence $(Q_iP_k-Q_kP_i)
(Q_iP_k-Q_kP_i)=2\{Q^2 P^2 - (Q\cdot P)^2\}$ must be
divisible by 4. Since $Q^2$ and $P^2$ are even, we must have
$(Q\cdot P)^2$ even and hence $Q\cdot P$ even. 
Thus the right hand side of \refb{ebb9} 
is divisible by $2\ell_1$. 
Furthermore for $\ell_1$ even
$\ell_2$ must be odd since $\ell_1$ and $\ell_2$ cannot
have a common factor. In this case
$\gcd(2\ell_1,\ell_2)=1$, and
we again conclude that $2\ell_1\ell_2$ divides \refb{ebb9}
since $2\ell_1$ and $\ell_2$ separately divides \refb{ebb9}. Thus
in either case we see that \refb{ebb9} is divisible by $2\ell_1\ell_2$
and hence by $2s$. This shows that \refb{ebb8} implies 
$2s|\chi(q)$
and we can express \refb{ebb4} as
\be \label{ebb10}
d(q)=(-1)^{Q\cdot P+1}
\sum_{s\in\zzz, s|\ell_1\ell_2} s\, \wh c(\Delta(Q,P)/s^2)\, .
\ee
This is the relation given in \refb{e3}.



\begin{thebibliography}{99}



\bibitem{9903163}
J.~Maldacena,  G.~Moore and A.~Strominger,
``Counting BPS blackholes in toroidal type II string theory,''
arXiv:hep-th/9903163.

\bibitem{0506151}
  D.~Shih, A.~Strominger and X.~Yin,
  ``Counting dyons in N = 8 string theory,''
  JHEP {\bf 0606}, 037 (2006)
  [arXiv:hep-th/0506151].

\bibitem{0506228}
  B.~Pioline,
  ``BPS black hole degeneracies and minimal automorphic representations,''
  JHEP {\bf 0508}, 071 (2005)
  [arXiv:hep-th/0506228].

\bibitem{0508174}
  D.~Shih and X.~Yin,
  ``Exact Black Hole Degeneracies and the Topological String,''
  JHEP {\bf 0604}, 034 (2006)
  [arXiv:hep-th/0508174].

\bibitem{0803.1014}
 A.~Sen,
  ``N=8 Dyon Partition Function and Walls of Marginal Stability,''
 JHEP {\bf 0807}, 118 (2008)
  [arXiv:0803.1014 [hep-th]].

\bibitem{0804.0651}
  A.~Sen,
  ``U-duality Invariant Dyon Spectrum in type II on $T^6$,''
 JHEP {\bf 0808}, 037 (2008)
  [arXiv:0804.0651 [hep-th]].

\bibitem{0805.0095}
  A.~Sen,
  ``Entropy Function and AdS(2)/CFT(1) Correspondence,''
  JHEP {\bf 0811}, 075 (2008)
  [arXiv:0805.0095 [hep-th]].

\bibitem{0809.3304}
  A.~Sen,
  ``Quantum Entropy Function from AdS(2)/CFT(1) Correspondence,''
  arXiv:0809.3304 [hep-th].

\bibitem{0903.1477}
  A.~Sen,
  ``Arithmetic of Quantum Entropy Function,''
  arXiv:0903.1477 [hep-th].

\bibitem{0903.2481}
  A.~Dabholkar, M.~Guica, S.~Murthy and S.~Nampuri,
  ``No entropy enigmas for N=4 dyons,''
  arXiv:0903.2481 [hep-th].

\bibitem{9708062}
  A.~Gregori, E.~Kiritsis, C.~Kounnas, N.~A.~Obers, 
  P.~M.~Petropoulos and B.~Pioline,
  ``R**2 corrections and non-perturbative 
  dualities of N = 4 string ground
  states,''
  Nucl.\ Phys.\ B {\bf 510}, 423 (1998)
  [arXiv:hep-th/9708062].

\bibitem{9708130}
  E.~Kiritsis,
  ``Introduction to non-perturbative string theory,''
  arXiv:hep-th/9708130.


\bibitem{cremmer}
  E.~Cremmer and B.~Julia,
  ``The SO(8) Supergravity,''
  Nucl.\ Phys.\  B {\bf 159}, 141 (1979).


\bibitem{9602014}
  R.~Kallosh and B.~Kol,
  ``E(7) Symmetric Area of the Black Hole Horizon,''
  Phys.\ Rev.\  D {\bf 53}, 5344 (1996)
  [arXiv:hep-th/9602014].

\bibitem{0810.3472}
  N.~Banerjee, D.~P.~Jatkar and A.~Sen,
  ``Asymptotic Expansion of the N=4 Dyon Degeneracy,''
  JHEP {\bf 0905}, 121 (2009)
  [arXiv:0810.3472 [hep-th]].


\bibitem{0904.4253}
  S.~Murthy and B.~Pioline,
  ``A Farey tale for N=4 dyons,''
  arXiv:0904.4253 [hep-th].

\bibitem{9508064}
  A.~Sen and C.~Vafa,
  ``Dual pairs of type II string compactification,''
  Nucl.\ Phys.\  B {\bf 455}, 165 (1995)
  [arXiv:hep-th/9508064].

\bibitem{9807056}
  G.~W.~Moore,
  ``Attractors and arithmetic,''
  arXiv:hep-th/9807056.


\bibitem{9807087}
  G.~W.~Moore,
  ``Arithmetic and attractors,''
  arXiv:hep-th/9807087.


\bibitem{0401049}
  G.~W.~Moore,
 ``Les Houches lectures on strings and arithmetic,''
  arXiv:hep-th/0401049.

\bibitem{0801.0149}
  S.~Banerjee and A.~Sen,
  ``S-duality Action on Discrete T-duality Invariants,''
  JHEP {\bf 0804}, 012 (2008)
  [arXiv:0801.0149 [hep-th]].


\bibitem{0712.0043}
S.~Banerjee and A.~Sen, 
``Duality Orbits, Dyon Spectrum and Gauge Theory Limit of
Heterotic String Theory on $T^6$'',
JHEP {\bf 0803}, 022 (2008)
  [arXiv:0712.0043 [hep-th]].

\bibitem{0905.2686}
  N.~Banerjee, S.~Banerjee, R.~Gupta, I.~Mandal and A.~Sen,
 ``Supersymmetry, Localization and Quantum Entropy Function,''
  arXiv:0905.2686 [hep-th].

\bibitem{0903.5517}
  L.~Borsten, D.~Dahanayake, M.~J.~Duff and W.~Rubens,
  ``Black holes admitting a Freudenthal dual,''
  arXiv:0903.5517 [hep-th].


 \end{thebibliography}
\end{document}